\documentclass[pra,twocolumn,showpacs,preprintnumbers,superscriptaddress]
{revtex4}

\usepackage{times}
\usepackage{bm}
\usepackage{graphicx}
\usepackage{amsbsy}
\usepackage{amsmath}
\usepackage{amsfonts}
\usepackage{amsthm}

\begin{document}

\theoremstyle{plain}
\newtheorem{theorem}{Theorem}
\newtheorem{lemma}[theorem]{Lemma}
\newtheorem{corollary}[theorem]{Corollary}
\newtheorem{proposition}[theorem]{Proposition}
\newtheorem{conjecture}[theorem]{Conjecture}

\theoremstyle{definition}
\newtheorem{definition}[theorem]{Definition}

\theoremstyle{remark}
\newtheorem*{remark}{Remark}
\newtheorem{example}{Example}
\title{Structural Physical Approximation make possible to realize the optimal singlet fraction with two measurements}
\author{Satyabrata Adhikari}
\email{satyabrata@dtu.ac.in} \affiliation{Delhi Technological
University, Delhi-110042, Delhi, India}

\begin{abstract}
Structural physical approximation (SPA) has been exploited to
approximate non-physical operation such as partial transpose. It
has already been studied in the context of detection of
entanglement and found that if the minimum eigenvalue of SPA to
partial transpose is less than $\frac{2}{9}$ then the two-qubit
state is entangled. We find application of SPA to partial
transpose in the estimation of optimal singlet fraction. We show
that optimal singlet fraction can be expressed in terms of minimum
eigenvalue of SPA to partial transpose. We also show that optimal
singlet fraction can be realized using Hong-Ou-Mandel
interferometry with only two detectors. Further we have shown that
the generated hybrid entangled state between a qubit and a binary
coherent state can be used as a resource state in quantum
teleportation.
\end{abstract}
 \maketitle

\section{Introduction}
Entanglement is a non-classical correlation \cite{piani} and is a
necessary ingredient to build a quantum computer that can
outperform the classical computer. It has also been used as a
quantum resource in various quantum communication tasks such as
teleportation \cite{Bennett1}, superdense coding \cite{Bennett2},
secret sharing \cite{hillery} and quantum-key distribution (QKD)
\cite{Gisin}. To perform the quantum communication task, we need
to generate entanglement but the issue in the generation of
entanglement is that it never free from imperfections and noise.
Therefore, the generated state may or may not be entangled. Thus
it is important to develop an efficient way to detect
entanglement.\\
Positive maps are stronger detectors of entanglement but they are
not physical and hence cannot be realized in the laboratory. On
the other hand, completely positive maps play an important role in
quantum information processing as they are capable to describe an
arbitrary quantum transmission channel \cite{schumacher}. Not only
that but also it approximated non-physical operations such as
quantum cloners or universal-NOT gate \cite{fiurasek1}. A physical
way by which positive maps can be approximated by completely
positive maps is called structural physical approximation (SPA)
\cite{horodecki2}. SPA thus transform non-physical operations to
physical operations.\\
Here, we restrict our discussion only to the two-qubit system. Let
$\textbf{\textsl{P}}$ denote the positive map in 4-dimensional
Hilbert space and $\textbf{\textsl{CP}}$ denote the completely
positive map that transforms all quantum states $\rho$ onto
maximally mixed state $\frac{I}{4}$ i.e.
$\textbf{\textsl{CP}}(\rho)=\frac{I}{4}$ \cite{fiurasek,korbicz}.
Since the positive map $\textbf{\textsl{P}}$ is not physically
realizable map so we approximate it with $\textbf{\textsl{CP}}$
map in such a way that it would be a physical map. Therefore, SPA
to the map $\textbf{\textsl{P}}$ is given by
\begin{eqnarray}
\widetilde{\textbf{\textsl{P}}}=(1-p^{*})\textbf{\textsl{P}}+\frac{p^{*}}{4}I_{4}
\label{spa}
\end{eqnarray}
where $p^{*}$ is the minimum value of $p$ for which
the approximated map $\widetilde{\textbf{\textsl{P}}}$ is a completely positive map \cite{bae}.\\
Partial transposition (PT) is another strong entanglement
detection criterion given by Peres \cite{peres}. Later, Horodecki
\cite{horodecki1} proved that the PT criterion is necessary and
sufficient for $2\times 2$ and $2\times 3$ system. Although
partial transposition criterion works well in qubit-qubit and
qubit-qutrit system but it cannot be implemented in a laboratory
for the detection of entanglement as it is a non-physical
operation. Therefore, to make partial transposition map a physical
operation we can approximate it in such a way that it would be a
completely positive map. Let us consider that the partial
transposition operation act on the second subsystem and is given
by $id\otimes\textbf{\textsl{T}}$, where $\textbf{\textsl{T}}$
denotes the positive transposition map and $id$ represent the
identity operator. In general, partial transposition operation
$id\otimes\textbf{\textsl{T}}$ can be approximated as
\cite{horodecki2}
\begin{eqnarray}
\widetilde{id\otimes\textbf{\textsl{T}}}=(1-q^{*})(id\otimes\textbf{\textsl{T}})+\frac{q^{*}}{4}I_{A}\otimes
I_{B} \label{spapt}
\end{eqnarray}
where $q^{*}=\frac{16\nu}{1+16\nu}$ and
$\nu=-min_{Q>0}Tr[Q(id\otimes\textbf{\textsl{T}})|\psi^{+}\rangle\langle\psi^{+}|]$,
$|\psi^{+}\rangle=\frac{1}{\sqrt{2}}(|00\rangle+|11\rangle)$. The
map $id\otimes\textbf{\textsl{T}}$  is a non-physical map but its
approximate map $\widetilde{id\otimes\textbf{\textsl{T}}}$ is a
completely positive map corresponds to a quantum channel that can
be experimentally implementable \cite{horodecki2,bae}.\\
Let $\sigma_{12}$ be a two qubit-state and the task is to
determine whether it is an entangled state or separable state.
Since it is a two qubit system so we can apply partial
transposition detection criterion. PT criterion states that if
$\sigma_{12}^{T_{B}}$ ($T_{B}$ denotes partial transposition with
respect to the second subsystem B) has a negative eigenvalue then
the state $\sigma_{12}$ is an entangled state. But partial
transposition is not a physical operation so apply SPA-PT
operation (\ref{spapt}) on $\sigma_{12}$ and at the output, we
have $\widetilde{\sigma_{12}}$. Since SPA-PT operation is
completely positive so the output $\widetilde{\sigma_{12}}$ also
represents a state. Therefore, the practical problem of finding
the eigenvalue of $\sigma_{12}^{T_{B}}$ ($T_{B}$ denotes partial
transposition with respect to the second subsystem B) reduces to
determine the eigenvalue of $\widetilde{\sigma_{12}}$. Hence PT
criterion modified as SPA-PT criterion which states that if the
minimum eigenvalue of $\widetilde{\sigma_{12}}$ is less than
$\frac{2}{9}$ then the state $\sigma_{12}$ is entangled and
vice-versa \cite{horodecki2}.
The minimum eigenvalue can be estimated by the procedure given in \cite{keyl,tanaka}.\\
Recently, H-T Lim et.al. \cite{lim} have demonstrated the
experimental realization of SPA-PT for photonic two qubit photonic
system using single-photon polarization qubits and linear optical
devices. They provided the decomposition of SPA-PT for a two-qubit
state $\sigma_{12}$ as
\begin{eqnarray}
\widetilde{\sigma_{12}}=\widetilde{I\otimes
T}(\sigma_{12})=[\frac{1}{3}(I\otimes\tilde{T})+\frac{2}{3}(\tilde{\Theta}\otimes
D)]\sigma_{12} \label{spa-pt decomposition}
\end{eqnarray}
where $\tilde{T}$ denote SPA for transpose operation, $\Theta$
denotes the inversion map and works as $\Theta(\sigma)=-\sigma$,
$\tilde{\Theta}$ denote its SPA and can be constructed by the
prescription given in (\ref{spa}) and $D(\sigma)=\frac{I_{2}}{2}$
denote the polarization. Since $I\otimes\tilde{T}$ and
$\tilde{\Theta}\otimes D$ are local operations and are completely
positive operators so $\widetilde{I\otimes T}$ is a physically realizable operators.\\
The motivation of this work is two fold: Firstly, the method of
finding the eigenvalues in \cite{keyl,tanaka} require more than
one copy of the given state and the method described in
\cite{keyl} for estimating the eigenvalues works well
asymtotically. Therefore, to circumvent these problems we take the
approach of witness operator to determine the minimum eigenvalue,
which require a single copy of SPA-PT of the given state. Also, we
show that the minimum eigenvalue determined by our method require
a set up that need only two measurements. Secondly, Verstraete and
Vershelde \cite{Verstraete} have established a relationship
between the optimal singlet fraction and partial transpose of a
given state and using the derived relation they have shown that
the two-qubit state is useful as a resource state for
teleportation if and only if the optimal singlet fraction is
greater than $\frac{1}{2}$. But the partial tranposition is an
non-physical operation and cannot be implemented in a laboratory
so it would be not easy to realize the optimal singlet fraction.
Also the filtering operation used in \cite{Verstraete} to achieve
the optimal singlet fraction depends on the quantum state under
investigation. Thus information about the state under
investigation is needed. To overcome these problems, we apply
SPA-PT method and show that optimal singlet fraction does not
depend on the state under investigation and also can be realized in experiment.\\
This paper is organized as follows: In section-II, we have
constructed the witness operator to determine the minimum
eigenvalue of SPA-PT of a given state $\rho_{12}$. In section-III,
we have shown that the number of measurements needed to determine
the minimum eigenvalue is two. In section-IV, we show that the
teleportation fidelity can be determined experimentally using
Hong-Ou-Mandel interferometry with only two detectors. In
section-V, we have studied the hybrid entangled state between a
qubit and binary coherent state and have shown that the mixed
hybrid entangled state can be used as a resource state for
teleportation and lastly, we conclude in section-VI.
\section{Witness operator that determine the minimum eigenvalue of SPA-PT of two qubit state}
Any arbitrary two qubit density operator in the
computational basis is given by
\begin{eqnarray}
\rho_{12}=
\begin{pmatrix}
  t_{11} & t_{12} & t_{13} & t_{14} \\
  t_{12}^{*} & t_{22} & t_{23} & t_{24} \\
  t_{13}^{*} & t_{23}^{*} & t_{33} & t_{34} \\
  t_{14}^{*} & t_{24}^{*} & t_{34}^{*} & t_{44}
\end{pmatrix}, \sum_{i=1}^{4}t_{ii}=1
\end{eqnarray}
where $(*)$ denotes the complex conjugate.\\\\
SPA-PT of $\rho_{12}$ is given by
\begin{eqnarray}
\widetilde{\rho_{12}}&=&[\frac{1}{3}(I\otimes\widetilde{T})+\frac{2}{3}(\widetilde{\Theta}\otimes\widetilde{D})]\rho_{12}\nonumber\\&=&
\begin{pmatrix}
  E_{11} & E_{12} & E_{13} & E_{14} \\
  E_{12}^{*} & E_{22} & E_{23} & E_{24} \\
  E_{13}^{*} & E_{23}^{*} & E_{33} & E_{34} \\
  E_{14}^{*} & E_{24}^{*} & E_{34}^{*} & E_{44}
\end{pmatrix}
\label{spa1}
\end{eqnarray}
where
\begin{eqnarray}
&&E_{11}=\frac{1}{9}(2+t_{11}),E_{12}=\frac{1}{9}(-it_{12}+t_{12}^{*}),\nonumber\\&&
E_{13}=\frac{1}{9}(t_{13}-i(t_{13}^{*}+t_{24}^{*})),
E_{14}=\frac{1}{9}(-it_{14}+t_{23}),\nonumber\\&&
E_{22}=\frac{1}{9}(2+t_{22}),E_{23}=\frac{1}{9}(t_{14}+it_{23}),\nonumber\\&&
E_{24}=\frac{-i}{9}(t_{13}^{*}+t_{24}^{*}),E_{33}=\frac{1}{9}(2+t_{33}),\nonumber\\&&
E_{34}=\frac{1}{9}(-it_{34}+t_{34}^{*}),E_{44}=\frac{1}{9}(2+t_{44})
\label{spa2a}
\end{eqnarray}
We note that the matrix $\widetilde{\rho_{12}}$ is not only a
Hermitian matrix but also has non-negative eigenvalues. The trace
of the matrix is equal to unity. So, it possesses all the
properties of a state and thus the matrix can be regarded as a
density matrix $\widetilde{\rho_{12}}$. The minimum eigenvalue of
$\widetilde{\rho_{12}}$ detect whether the state $\rho_{12}$ is
entangled or not? Therefore, our task is to construct the witness
operator that detect whether the minimum eigenvalue of
$\widetilde{\rho_{12}}$ is less than $\frac{2}{9}$?\\
To start, we consider the operator
$\textsl{\textit{O}}=\widetilde{\rho_{12}}-\frac{1}{9}\rho_{12}^{T_{2}}$,
$T_{2}$ denote the partial transpose with respect to the second
subsystem. The expectation value of the operator
$\textsl{\textit{O}}$ in the state
$|\phi\rangle=\alpha|00\rangle+\beta|11\rangle(
\alpha^{2}+\beta^{2}=1)$ is given by
\begin{eqnarray}
&&\langle \phi|O|\phi\rangle=
Tr[(\widetilde{\rho_{12}}-\frac{1}{9}\rho_{12}^{T_{2}})|\phi\rangle\langle\phi|]=\frac{2}{9}\nonumber\\&&
\Rightarrow
Tr[|\phi\rangle\langle\phi|\widetilde{\rho_{12}}]-\frac{1}{9}Tr[|\phi\rangle\langle\phi|^{T_{2}}\rho_{12}]=\frac{2}{9}\nonumber\\&&
\Rightarrow
Tr[W^{opt}\rho_{12}]=Tr[\widetilde{\rho_{12}}|\phi\rangle\langle\phi|]-\frac{2}{9}
\label{expectationvalue}
\end{eqnarray}
where $W^{opt}=\frac{1}{9}|\phi\rangle\langle\phi|^{T_{2}}$ is the
optimal witness operator that detect whether the state $\rho_{12}$
is entangled or not.\\
Let $\lambda_{min}$ be the minimum eigenvalue of
$\widetilde{\rho_{12}}$ and $|\phi\rangle$ be the eigenvector
corresponding to the minimum eigenvalue, then the eigenvalue
equation is given by
\begin{eqnarray}
\widetilde{\rho_{12}}|\phi\rangle=\lambda_{min}|\phi\rangle
\label{eigenvalueequ}
\end{eqnarray}
Using (\ref{eigenvalueequ}) in (\ref{expectationvalue}), we have
\begin{eqnarray}
Tr(W^{opt}\rho_{12})=\lambda_{min}-\frac{2}{9}
\label{expectationeigen}
\end{eqnarray}
For all separable state $\rho_{12}^{s}$, we have \cite{horodecki2}
\begin{eqnarray}
Tr(W^{opt}\rho_{12}^{s})\geq0\Rightarrow
\lambda_{min}\geq\frac{2}{9} \label{condsepeigen}
\end{eqnarray}
If $\rho_{12}$ is an entangled state and $W^{opt}$ detect that
entangled state then
\begin{eqnarray}
Tr(W^{opt}\rho_{12})<0\Rightarrow \lambda_{min}<\frac{2}{9}
\label{condenteigen}
\end{eqnarray}
The inequality (\ref{condenteigen}) gives us the condition that
when $\rho_{12}$ is an entangled state.\\
Since the above condition is a purely mathematical condition so
naturally one can ask a question that can we achieve this
inequality experimentally? To investigate this, let us again
recall (\ref{expectationvalue}) and write it in a different form
as
\begin{eqnarray}
Tr(W^{opt}\rho_{12})=Tr[(|\phi\rangle\langle\phi|-\frac{2}{9}I)\widetilde{\rho_{12}}]
\label{expectationvalueoperator}
\end{eqnarray}
When $W^{opt}$ detect an entangled state $\rho_{12}$ then
$Tr(W^{opt}\rho_{12})<0$ and hence we arrive at a condition given
by
\begin{eqnarray}
Tr[(|\phi\rangle\langle\phi|-\frac{2}{9}I)\widetilde{\rho_{12}}]<0
\label{detectioncond}
\end{eqnarray}
The above condition (\ref{detectioncond}) is equivalent form of
the condition (\ref{condenteigen}) and hence the inequality
implies that the eigenvalues of $\widetilde{\rho_{12}}$ is less
than $\frac{2}{9}$.\\
Let $\textit{\textbf{V}}\equiv
|\phi\rangle\langle\phi|-\frac{2}{9}I$. Then the inequality
(\ref{detectioncond}) can be re-expressed as
\begin{eqnarray}
Tr(\textit{\textbf{V}}\widetilde{\rho_{12}})<0
\label{detectioncond1}
\end{eqnarray}
Next we investigate few properties of the operator $\textit{\textbf{V}}$.\\
\textbf{P1.}The expectation value of $\textit{\textbf{V}}$ for
all separable state $\widetilde{\rho_{12}}^{sep}$ is non-negative i.e. $Tr(\textit{\textbf{V}}\widetilde{\rho_{12}}^{sep})\geq 0$. \\
Proof: The  separable state $\widetilde{\rho_{12}}^{sep}$ is given
by
\begin{eqnarray}
\widetilde{\rho_{12}}^{sep}=
\begin{pmatrix}
  E_{11}^{s} & E_{12}^{s} & E_{13}^{s} & E_{14}^{s} \\
  (E_{12}^{s})^{*} & E_{22}^{s} & E_{23} & E_{24} \\
  (E_{13}^{s})^{*} & (E_{23}^{s})^{*} & E_{33}^{s} & E_{34}^{s} \\
  (E_{14}^{s})^{*} & (E_{24}^{s})^{*} & (E_{34}^{s})^{*}& E_{44}^{s}
\end{pmatrix}
\label{spasep}
\end{eqnarray}
where
$E_{11}^{s}=\frac{1}{9}(2+t_{11}^{s}),E_{12}^{s}=\frac{1}{9}(-it_{12}^{s}+(t_{12}^{s})^{*}),E_{13}^{s}=\frac{1}{9}(t_{13}^{s}-i((t_{13}^{s})^{*}+(t_{24}^{s})^{*}),
E_{14}=\frac{1}{9}(-it_{14}^{s}+t_{23}^{s}),E_{22}^{s}=\frac{1}{9}(2+t_{22}^{s}),E_{23}^{s}=\frac{1}{9}(t_{14}^{s}+it_{23}^{s}),
E_{24}^{s}=\frac{-i}{9}((t_{13}^{s})^{*}+(t_{24}^{s})^{*}),
E_{33}^{s}=\frac{1}{9}(2+t_{33}^{s}),
E_{34}^{s}=\frac{1}{9}(-it_{34}^{s}+(t_{34}^{s})^{*}),E_{44}^{s}=\frac{1}{9}(2+t_{44}^{s})$.
Here, $t_{ij}^{s}$ denote the elements of a separable state $\rho_{12}^{s}$.\\
(\ref{expectationvalueoperator}) can be re-expressed for any
separable state $\rho_{12}^{s}$ as
\begin{eqnarray}
Tr(W^{opt}\widetilde{\rho_{12}}^{sep})=Tr(\textit{\textbf{V}}\widetilde{\rho_{12}}^{sep})
\label{deduct3}
\end{eqnarray}
Since $W^{opt}$ is a witness operator so the expectation value of
$W^{opt}$ over all separable state $\widetilde{\rho_{12}}^{sep}$
is non-negative. Thus
\begin{eqnarray}
 Tr(W^{opt}\widetilde{\rho_{12}}^{sep})\geq 0
 \label{deduct4}
\end{eqnarray}
Using (\ref{deduct3}) and (\ref{deduct4}), we have
\begin{eqnarray}
Tr(\textit{\textbf{V}}\widetilde{\rho_{12}}^{sep})\geq 0
\label{deduct5}
\end{eqnarray}
\textbf{P2.} It can be easily shown that $\textit{\textbf{V}}$ has
at least one negative eigenvalues.\\
Thus, the operator $\textit{\textbf{V}}$ possess all the
properties of a witness operator and hence it detect whether the
eigenvalue of $\widetilde{\rho_{12}}$ is less than $\frac{2}{9}$
or not. If $\textit{\textbf{V}}$ detect that the eigenvalue of
$\widetilde{\rho_{12}}$ is less than $\frac{2}{9}$ then we can say
that the state described by the density operator $\rho_{12}$ is
entangled.\\
Since $\textit{\textbf{V}}$ is a hermitian operator so it is an
observable and can be implemented experimentally. Therefore the
inequality (\ref{detectioncond}) is useful to detect entangled
state experimentally.
\section{Number of measurements needed to determine the value of $Tr(\textit{\textbf{V}}\widetilde{\rho_{12}})$}
The operator $\textit{\textbf{V}}$ can be expressed in terms of
local Pauli matrices as
\begin{eqnarray}
\textit{\textbf{V}}&=&\frac{9}{28}[\frac{7}{9}I\otimes
I+(\alpha^{2}-\beta^{2})(I\otimes \sigma_{z}+\sigma_{z} \otimes
I)\nonumber\\&+&2\alpha\beta(\sigma_{x}\otimes\sigma_{x}+\sigma_{y}\otimes\sigma_{y})+\sigma_{z}\otimes\sigma_{z}]
 \label{paulimatrices}
\end{eqnarray}
We find that the decomposition of the operator
$\textit{\textbf{V}}$ in terms of local Pauli observables need
more than two measurements to realize it. So in this section, our
task is to show that it is possible to realize the operator
$\textit{\textbf{V}}$ with just two measurements.\\
To achieve our goal, we approximate the entanglement witness
operator $\textit{\textbf{V}}$ in a way we approximate the
positive but not completely positive operator. Therefore,
approximate entanglement witness operator
$\widetilde{\textit{\textbf{V}}}$ can be expressed as
\begin{eqnarray}
\widetilde{\textit{\textbf{V}}}=p \textit{\textbf{V}}+(1-p)I,
0\leq p\leq 1 \label{aew}
\end{eqnarray}
Choose the minimum value of $p$ in such a way that the operator
$\widetilde{\textit{\textbf{V}}}$ will become a positive
semi-definite operator. The Hermitian operator
$\widetilde{\textit{\textbf{V}}}$ is positive semi-definite if
$p_{min}=\frac{8}{15}$. Therefore,
$\widetilde{\textit{\textbf{V}}}$ can be re-expressed as
\begin{eqnarray}
\widetilde{\textit{\textbf{V}}}=\frac{8}{15}\textit{\textbf{V}}+\frac{7}{15}I,
\label{aew1}
\end{eqnarray}
We can observe that the operator $\widetilde{\textit{\textbf{V}}}$
is not a normalized operator so it can be expressed after
normalization as
\begin{eqnarray}
\widetilde{\textit{\textbf{V}}}=\frac{2}{9}\textit{\textbf{V}}+\frac{7}{36}I,
\label{aew1}
\end{eqnarray}
Since $\widetilde{\textit{\textbf{V}}}$ is positive semi-definite
and $Tr(\widetilde{\textit{\textbf{V}}})=1$ so the operator
$\widetilde{\textit{\textbf{V}}}$ can be treated as a quantum
state.\\
Again, $Tr(\textit{\textbf{V}}\widetilde{\rho_{12}})$ can be
written in terms of
$Tr(\widetilde{\textit{\textbf{V}}}\widetilde{\rho_{12}})$ as
\begin{eqnarray}
Tr(\textit{\textbf{V}}\widetilde{\rho_{12}})=\frac{15}{8}Tr(\widetilde{\textit{\textbf{V}}}\widetilde{\rho_{12}})-\frac{7}{8}
\label{exp1}
\end{eqnarray}
It can be shown that
$Tr(\widetilde{\textit{\textbf{V}}}\widetilde{\rho_{12}})$ is
equal to the average fidelity for two mixed quantum states
$\widetilde{\textit{\textbf{V}}}$ and $\widetilde{\rho_{12}}$
\cite{kwong},
\begin{eqnarray}
Tr(\widetilde{\textit{\textbf{V}}}\widetilde{\rho_{12}})=F_{avg}(\widetilde{\textit{\textbf{V}}},\widetilde{\rho_{12}})
\label{avfid}
\end{eqnarray}
Using (\ref{exp1}) and (\ref{avfid}), we have
\begin{eqnarray}
Tr(\textit{\textbf{V}}\widetilde{\rho_{12}})=\frac{15}{8}F_{avg}(\widetilde{\textit{\textbf{V}}},\widetilde{\rho_{12}})-\frac{7}{8}
\label{exp2}
\end{eqnarray}
C. J. Kwong et.al. \cite{kwong} have shown that the average
fidelity between two mixed quantum states can be estimated
experimentally by Hong-Ou-Mandel interferometry with only two
detectors. Thus the quantity
$Tr(\textit{\textbf{V}}\widetilde{\rho_{12}})$
needs only two measurements to estimate it in experiment.\\
Again, the minimum eigenvalue can be expressed in terms of
$F_{avg}(\widetilde{\textit{\textbf{V}}},\widetilde{\rho_{12}})$
as
\begin{eqnarray}
\lambda_{min}=\frac{15}{8}F_{avg}(\widetilde{\textit{\textbf{V}}},\widetilde{\rho_{12}})-\frac{47}{72}
\label{mineigen}
\end{eqnarray}
Since the minimum eigenvalue of the quantum state
$\widetilde{\rho_{12}}$ can be determined using two measurements
and minimum eigenvalue is responsible for the detection of
entanglement so we can say that the presence of entanglement in
$\rho_{12}$ can be detected using two measurements only.
\section{Realization of optimal singlet fraction with only two measurements}
The singlet fraction is defined as
\begin{eqnarray}
F(\rho_{12})=max[
\langle\phi^{+}|\rho_{12}|\phi^{+}\rangle\,\langle\phi^{-}|\rho_{12}|\phi^{-}\rangle\,\\
\langle\psi^{+}|\rho_{12}|\psi^{+}\rangle\,\langle\psi^{-}|\rho_{12}|\psi^{-}\rangle]
\label{singletfrac}
\end{eqnarray}
where
$\{|\phi^{+}\rangle,|\phi^{-}\rangle,|\psi^{+}\rangle,|\psi^{-}\rangle\}$
are the maximally entangled Bell states.\\
Verstraete and Vershelde \cite{Verstraete} suggested the optimal
trace preserving protocol for maximizing the singlet fraction of a
given state. The optimal singlet fraction is given by
\begin{eqnarray}
F^{opt}(\rho_{12})=\frac{1}{2}-Tr(X_{opt}\rho_{12}^{T_{B}})
\label{optsingletfrac}
\end{eqnarray}
$X_{opt}$ is given by
\begin{eqnarray}
X_{opt}=(A\otimes
I_{2})|\psi^{+}\rangle\langle\psi^{+}|(A^{\dagger}\otimes I_{2}),
\label{filter}
\end{eqnarray}
where $I_{2}$ represent an identity matrix of order 2,
$|\psi^{+}\rangle=\frac{1}{\sqrt{2}}(|00\rangle+|11\rangle)$ and
$A=\begin{pmatrix}
  a & 0 \\
  0 & 1
\end{pmatrix}, -1\leq a \leq1$.\\
We note that if the state $\rho_{12}$ is entangled then
$\rho_{12}^{T_{B}}$ has at least one negative eigenvalue and hence
$F^{opt}(\rho_{12})$ is greater than $\frac{1}{2}$. Thus every
entangled two qubit state is useful for teleportation. But there
are two problems in estimating the quantity $F^{opt}(\rho_{12})$
given in (\ref{optsingletfrac}): (i)$\rho_{12}^{T_{B}}$ cannot be
realized in the laboratory and (ii) the parameter in $X_{opt}$ is
state dependent and hence to construct $X_{opt}$, we need to know
the state under investigation.\\
In this section, we obtain the optimal singlet fraction in terms
of minimum eigenvalue $\lambda_{min}$ of SPA-PT of the state
$\rho_{12}$ using local filtering operation but in this scenario
the parameter of the local filtering operation does not depend on
the state under investigation. Since we found in the previous
section that $\lambda_{min}$ can be estimated for any arbitrary
state with two measurements so we can say that optimal singlet
fraction of any arbitrary states can be calculated using two
measurements.
\subsection{Optimal singlet fraction in terms of $\lambda_{min}$}
To start, let us consider the operator
$X_{opt}(\widetilde{\rho_{12}}-\frac{1}{9}\rho_{12}^{T_{2}})$,
where $X_{opt}$ is given in (\ref{filter}). Calculate the trace of
it and after some simple algebra, (\ref{optsingletfrac}) reduces
to
\begin{eqnarray}
F^{opt}(\rho_{12})=\frac{1}{2}-[9Tr(X_{opt}\widetilde{\rho_{12}})-(a^{2}+1)]
\label{approxsingletfrac}
\end{eqnarray}
It can be easily seen from (\ref{approxsingletfrac}) that
$F^{opt}(\rho_{12})>\frac{1}{2}$ if and only if
$9Tr(X_{opt}\widetilde{\rho_{12}})-(a^{2}+1)<0$. This condition
leads to
\begin{eqnarray}
Tr(X_{opt}\widetilde{\rho_{12}})<\frac{a^{2}+1}{9}, -1\leq a\leq 1
\label{teleportcond}
\end{eqnarray}
In $\{|00\rangle,|11\rangle\}$ subspace, the operator
$\frac{1}{a^{2}+1}X$ can be expressed as
\begin{eqnarray}
\frac{2}{a^{2}+1}X_{opt}=|\chi\rangle\langle\chi| \label{state}
\end{eqnarray}
where
$|\chi\rangle=\frac{1}{\sqrt{a^{2}+1}}(a|00\rangle+|11\rangle)$.\\
We note that the vector $|\chi\rangle$ and the eigenvector
$|\phi\rangle$ of the operator $\widetilde{\rho_{12}}$ are
parallel vectors and thus there exist a real scalar $k$ such that
\begin{eqnarray}
|\chi\rangle=k |\phi\rangle \label{parallelvec}
\end{eqnarray}
Using (\ref{parallelvec}), it can be easily shown that the vector
$|\chi\rangle$ is also a eigenvector corresponding to the minimum
eigenvalue $\lambda_{min}$. Hence, the resource state $\rho_{12}$
is useful for teleportation iff
\begin{eqnarray}
\lambda_{min}<\frac{2}{9} \label{teleeigenvalcond}
\end{eqnarray}
The singlet fraction $F^{opt}(\rho_{12})$ given by
(\ref{approxsingletfrac}) can be re-expressed in terms of the
minimum eigenvalue $\lambda_{min}$ as
\begin{eqnarray}
F^{opt}_{(a,\lambda_{min})}(\rho_{12})&=&\frac{1}{2}-\frac{9(a^{2}+1)}{2}[\lambda_{min}-\frac{2}{9}],\nonumber\\&&
-1\leq a \leq 1 \label{singfraceigenvalcond}
\end{eqnarray}
Further, we find that if the minimum eigenvalue $\lambda_{min}$ is
restricted to lie in the interval $[\frac{1}{6},\frac{2}{9})$ then
the singlet fraction $F^{opt}_{(a,\lambda_{min})}(\rho_{12})$ lies
in the interval
\begin{eqnarray}
\frac{1}{2}<F^{opt}_{(a,\lambda_{min})}(\rho_{12})<\frac{1}{2}+\frac{a^{2}+1}{4},
-1\leq a \leq 1 \label{singfracrange}
\end{eqnarray}
The optimal singlet fraction can be achieved by putting $a=\pm1$
in (\ref{singfraceigenvalcond}) and it is given by
\begin{eqnarray}
F^{opt}_{(\pm1,\lambda_{min})}(\rho_{12})=\frac{1}{2}-9[\lambda_{min}-\frac{2}{9}],
\frac{1}{6}\leq \lambda_{min}<\frac{2}{9} \label{optsingfrac}
\end{eqnarray}
Without any loss of generality, we can take $a=1$ and thus we have
$X^{opt}=|\psi^{+}\rangle\langle\psi^{+}|$. Afterward, we denote
$F^{opt}_{(\pm1,\lambda_{min})}(\rho_{12})$ as simply
$F^{opt}_{\lambda_{min}}(\rho_{12})$. Therefore,
\begin{eqnarray}
F^{opt}_{\lambda_{min}}(\rho_{12})=\frac{1}{2}-9[\lambda_{min}-\frac{2}{9}],
\frac{1}{6}\leq \lambda_{min}<\frac{2}{9} \label{optsingfrac1}
\end{eqnarray}
We note here that the problem of finding the optimal singlet
fraction reduces to finding the minimum eigenvalue of SPA-PT of
any arbitrary state $\rho_{12}$.
\subsection{Optimal singlet fraction in terms of average fidelity
$F_{avg}(\widetilde{\textit{\textbf{V}}},\widetilde{\rho_{12}})$}
Now, we are in a position to give the expression for optimal
singlet fraction in terms of average fidelity
$F_{avg}(\widetilde{\textit{\textbf{V}}},\widetilde{\rho_{12}})$.\\
The optimal singlet fraction in terms of $\lambda_{min}$ can be
re-written as
\begin{eqnarray}
F^{opt}_{\lambda_{min}}(\rho_{12})=\frac{1}{2}-9[\lambda_{min}-\frac{2}{9}],
\frac{1}{6}\leq \lambda_{min}<\frac{2}{9}
\label{optsingfraceigval}
\end{eqnarray}
Using (\ref{mineigen}) in (\ref{optsingfraceigval}), we get
\begin{eqnarray}
F^{opt}_{\lambda_{min}}(\rho_{12})&=&\frac{1}{2}-\frac{135}{8}[F_{avg}(\widetilde{\textit{\textbf{V}}},\widetilde{\rho_{12}})-\frac{7}{15}],\nonumber\\&&
\frac{59}{135}\leq
F_{avg}(\widetilde{\textit{\textbf{V}}},\widetilde{\rho_{12}})<\frac{7}{15}
\label{optsingfracavgfid}
\end{eqnarray}
Since
$F_{avg}(\widetilde{\textit{\textbf{V}}},\widetilde{\rho_{12}})$
can be determined experimentally \cite{kwong} so
$F^{opt}_{\lambda_{min}}(\rho_{12})$ can be realized using
Hong-Ou-Mandel interferometry with only two detectors.
\subsection{Optimal Teleportation Fidelity for Two Qubit System}
For two qubit system, optimal teleportation fidelity
$f_{opt}(\rho_{12})$ and optimal singlet fraction
$F^{opt}_{\lambda_{min}}(\rho_{12})$ are related by
\cite{horodecki}
\begin{eqnarray}
f_{opt}(\rho_{12})&=&\frac{2F^{opt}_{\lambda_{min}}(\rho_{12})+1}{3}\nonumber\\&&
=\frac{2}{3}-\frac{135}{12}[F_{avg}(\widetilde{\textit{\textbf{V}}},\widetilde{\rho_{12}})-\frac{7}{15}]\nonumber\\&&
\frac{59}{135}\leq
F_{avg}(\widetilde{\textit{\textbf{V}}},\widetilde{\rho_{12}})<\frac{7}{15}
\label{telesingletfrac}
\end{eqnarray}
We can say a teleportation scheme is quantum if teleportation
fidelity is greater than $\frac{2}{3}$. We can find that the
teleportation fidelity given in (\ref{telesingletfrac}) is always
greater than $\frac{2}{3}$. Since $f_{opt}(\rho_{12})$ depends
only on
$F_{avg}(\widetilde{\textit{\textbf{V}}},\widetilde{\rho_{12}})$
so again the optimal teleportation fidelity can be realized by
Hong-Ou-Mandel interferometry with only two detectors.

\section{Application}
In this section, we will study a particular type of hybrid
entangled system prepared with qubit and binary coherent state
(BCS) in the context of quantum teleportation. Although coherent
state is described by infinite dimensional Hilbert space but
binary coherent state can be described by two dimensional Hilbert
space \cite{muller}. The states $|+\alpha\rangle$ and
$|-\alpha\rangle$ are called BCS and the set
$\{|+\alpha\rangle,|-\alpha\rangle\}$ forms a non-orthogonal BCS
basis. The state $|\alpha\rangle$ is given by
\begin{eqnarray}
|\alpha\rangle=exp(\frac{-|\alpha|^{2}}{2})\sum_{n}\frac{\alpha^{n}}{n!}|n\rangle
\label{coherentstate}
\end{eqnarray}
The BCS basis can be expressed in terms of computational basis as
\cite{muller}
\begin{eqnarray}
&&|+\alpha\rangle=cos(\theta)|0\rangle+sin(\theta)|1\rangle,
\nonumber\\&&
|-\alpha\rangle=sin(\theta)|0\rangle+cos(\theta)|1\rangle,~~~~~~0<\theta\leq\frac{\pi}{4}
\label{transformation}
\end{eqnarray}
The parameter $\theta$ can be determined by the overlapping
between the non-orthogonal states $|+\alpha\rangle$ and
$|-\alpha\rangle$ and it is given by
\begin{eqnarray}
\langle+\alpha|-\alpha\rangle=sin(2\theta) \label{orthogonalcond}
\end{eqnarray}
\subsection{Generation of Hybrid Entangled State between a qubit and BCS}
We now describe the method suggested in \cite{yurke} for the
generation of hybrid entangled state between a qubit and BCS. We
should note that strong Kerr nonlinear media can be used to
generate hybrid entanglement but this nonlinear effects in
existing media are extremely weak. Thus it would be better to use
weak Kerr nonlinearity to generate entanglement. Weak Kerr
nonlinearity interaction Hamiltonian is given by $H_{k}=\hbar \chi
a_{1}^{\dagger}a_{1}a_{2}^{\dagger}a_{2}$. The interaction between
a single-photon qubit
$|\psi\rangle_{1}=c|0\rangle_{1}+d|1\rangle_{1},
|c|^{2}+|d|^{2}=1$ and a coherent state $|\alpha\rangle_{2}$ under
interaction Hamiltonian $H_{k}$ is described as
\begin{eqnarray}
|\Psi^{(\vartheta)}\rangle_{12}&=&exp(\frac{iH_{k}t}{\hbar})|\psi\rangle_{1}|\alpha\rangle_{2}\nonumber\\&=&c|0\rangle_{1}|\alpha\rangle_{2}+d|1\rangle_{1}|\alpha
exp(i\vartheta)\rangle_{2} \label{interaction}
\end{eqnarray}
Taking $\vartheta=\pi$ in (\ref{interaction}), the state
$|\Psi^{(\vartheta)}\rangle_{12}$ reduces to
\begin{eqnarray}
|\Psi^{\vartheta=\pi}\rangle_{12}=c|0\rangle_{1}|\alpha\rangle_{2}+d|1\rangle_{1}|-\alpha
\rangle_{2} \label{hybrid}
\end{eqnarray}
(\ref{hybrid}) represent a hybrid entangled system between a qubit
and BCS.\\
We may note here that an optical fibre of about 3000 km is
required for $\vartheta=\pi$ for an optical frequency of
$\omega=5\times10^{14}$ rad/sec using currently available Kerr
nonlinearity \cite{sanders}.
\subsection{Qubit-BCS Hybrid Entangled State as a Non-Maximally
Two Qubit entangled State}

We will now show that the hybrid entangled state (\ref{hybrid})
can be used as a resource state for quantum teleportation. To
accomplish our task, we first express BCS in computational basis
as in (\ref{transformation}) and then treat the hybrid entangled
state as an entangled state in four dimensional Hilbert space.
Therefore, qubit-BCS hybrid entangled state (\ref{hybrid}) can be
expressed in the computational basis as
\begin{eqnarray}
|\Psi_{12}\rangle&=&c cos(\theta)|00\rangle_{12}+ c
sin(\theta)|01\rangle_{12}+ d
sin(\theta)|10\rangle_{12}\nonumber\\&& + d
cos(\theta)|11\rangle_{12} \label{hybrid1}
\end{eqnarray}
Let us assume that Alice have generated the hybrid entangled state
(\ref{hybrid1}). She attach an ancilla prepared in $|0\rangle_{a}$
to the state $|\Psi_{12}\rangle$. Therefore, the resulting three
qubit state $|\Psi_{12a}\rangle$ is given by
\begin{eqnarray}
|\Psi_{12a}\rangle&=&[c cos(\theta)|00\rangle_{12}+ c
sin(\theta)|01\rangle_{12}+ d
sin(\theta)|10\rangle_{12}\nonumber\\&& + d
cos(\theta)|11\rangle_{12}]\otimes |0\rangle_{a}
\label{hybridancilla}
\end{eqnarray}
Alice then apply two qubit CNOT-gate on qubit '2' and qubit 'a'.
The state $|\Psi_{12a}\rangle$ reduces to
\begin{eqnarray}
|\Phi_{12a}\rangle&=&c |0\rangle_{1}\otimes
(cos(\theta)|00\rangle_{2a}+
sin(\theta)|11\rangle_{2a})\nonumber\\&&+ d |1\rangle_{1}\otimes
(sin(\theta)|00\rangle_{2a}+ cos(\theta)|11\rangle_{2a})
\label{hybridancillacnot}
\end{eqnarray}
She performs a single qubit measurement in
$\{|0\rangle_{1},|1\rangle_{1}\}$ basis. If the measurement result
is $|0\rangle_{1}$ then with probability $|c|^{2}$, she prepare
the state
\begin{eqnarray}
|\Phi_{2a}^{(1)}\rangle=cos(\theta)|00\rangle_{2a}+
sin(\theta)|11\rangle_{2a} \label{hybridancillacnotresult0}
\end{eqnarray}
Again, if the measurement result is $|1\rangle_{1}$ then with
probability $|d|^{2}$, she prepare the state
\begin{eqnarray}
|\Phi_{2a}^{(2)}\rangle=sin(\theta)|00\rangle_{2a}+
cos(\theta)|11\rangle_{2a} \label{hybridancillacnotresult1}
\end{eqnarray}

\subsection{Mixed Qubit-BCS Entangled System Shared Between Two
Distant Parties As A Resource State For Quantum Teleportation}

\textbf{Case-I: }Let us assume that Alice have succeeded to
generate the non-maximally entangled state
$|\Phi_{2a}^{(1)}\rangle$ given by
(\ref{hybridancillacnotresult0}). She now want to share a
subsystem with her distant partner Bob so that they can use the
shared entangled state in sending the quantum information. To
achieve this, Alice send the subsystem '2' to Bob through the
memoryless amplitude damping channel. The transformation under
memoryless amplitude damping channel with parameter p $(0\leq
p\leq1)$ that governs the evolution of the system and the
environment is given by
\begin{eqnarray}
&&|0\rangle_{2}\otimes |0\rangle_{E}\rightarrow
|0\rangle_{2}\otimes |0\rangle_{E} \nonumber\\&&
|1\rangle_{2}\otimes |0\rangle_{E}\rightarrow
\sqrt{1-p}|1\rangle_{2}\otimes
|0\rangle_{E}+\sqrt{p}|0\rangle_{2}\otimes |1\rangle_{E}
\label{transformationampdamp}
\end{eqnarray}
When the qubit in mode '2' of the hybrid system
$|\Phi_{2a}^{1)}\rangle$ passes through the memoryless amplitude
damping channel then the system evolve as a mixed state and it is
given by
\begin{eqnarray}
&&\rho_{2a}=\sum_{i=0}^{1}(K_{i}\otimes
I)|\Phi_{2a}^{(1)}\rangle\langle\Phi_{2a}^{(1)}|(K_{i}^{\dagger}\otimes
I)\nonumber\\&&=
\begin{pmatrix}
cos^{2}(\theta) & 0 & 0 & \frac{\sqrt{1-p}}{2}sin(2\theta) \\
  0 & psin^{2}(\theta) & 0 & 0 \\
  0 & 0 & 0 & 0 \\
 \frac{\sqrt{1-p}}{2}sin(2\theta) & 0 & 0 & (1-p)sin^{2}(\theta)
\end{pmatrix}
\label{mixedstateampdampmemoless}
\end{eqnarray}
where the Kraus operators $K_{0}$ and $K_{1}$ are given by
\begin{eqnarray}
K_{0}=
\begin{pmatrix}
1 & 0 \\
  0 & \sqrt{1-p}
\end{pmatrix},K_{1}=
\begin{pmatrix}
1 & \sqrt{p} \\
  0 & 0
\end{pmatrix}
\label{kraus}
\end{eqnarray}
Now our task is to investigate whether the mixed state $\rho_{2a}$
shared between Alice and Bob is still entangled and if it is
entangled then under what condition? If we find that the state
$\rho_{2a}$ is entangled under certain conditions then the state
$\rho_{2a}$ can be used as a resource state for quantum
teleportation. To probe the above question, we calculate the
optimal singlet fraction given in (\ref{optsingfrac1}). The
optimal singlet fraction for the state $\rho_{2a}$ is given by
\begin{eqnarray}
F^{opt}_{(p,\theta)}(\rho_{2a})&=&\frac{1}{2}+\frac{1}{2}[\sqrt{(1-p)sin^{2}(2\theta)+p^{2}sin^{4}(\theta)}\nonumber\\&&-psin^{2}(\theta)],
0<\theta\leq \frac{\pi}{4} \label{optsingfracrho2a}
\end{eqnarray}
For $0<p<1$ and $0<\theta\leq\frac{\pi}{4}$, we have
\begin{eqnarray}
F^{opt}_{(p,\theta)}(\rho_{2a})>\frac{1}{2}
\label{optsingfracrho2a1}
\end{eqnarray}
Hence, the hybrid system described by the density matrix
$\rho_{2a}$ is useful as a resource state for the teleportation of
a single qubit. Further we observe the following points:\\
(i) For any value of the non-orthogonal parameter $\theta$
$(0<\theta<1)$, the value of the quantity
$F^{opt}_{(p,\theta)}(\rho_{2a})$ decreases as the noise parameter
$p$ increases from zero to unity.\\
(ii) For any value of the noise parameter $p$ $(0<p<1)$, the value
of the quantity $F^{opt}_{(p,\theta)}(\rho_{2a})$ increases as the
non-orthogonal parameter $\theta$ increases from zero to
$\frac{\pi}{4}$.\\\\
\textbf{Case-II: }If Alice have succeeded to generate the
non-maximally entangled state $|\Phi_{2a}^{(2)}\rangle$ given by
(\ref{hybridancillacnotresult1}) then also result remains the same
as in case-I.

\section{Conclusion}
To summarize, we have constructed the witness operator by
approximating the partial transposition operation, that can detect
the entangled state in a bipartite system. The constructed witness
operator can be decomposed into Pauli matrices and therefore, we
find that it need more than two measurements to realize the
witness operator. To minimize the number of measurements, we
approximate the entanglement witness, which is a positive
semi-definite operator. Then we express the minimum eigenvalue of
SPA-PT of the state under investigation in terms of average
fidelity between the approximated entanglement witness and SPA-PT
of the state. The latter one can be realized by Hong-Ou-Mandel
interferometry with only two detectors. So we infer that minimum
eigenvalue of SPA-PT of the state can be realized by
Hong-Ou-Mandel interferometry with only two detectors. Further, we
have obtained the minimum eigenvalue $\frac{1}{6}$ for a large
class of states and then we claim that it holds for any two qubit
states. We derived a relation between the optimal singlet fraction
and the minimum eigenvalue of SPA-PT of the state under
investigation. Therefore, optimal singlet fraction can also be
realized by Hong-Ou-Mandel interferometry with only two
detectors. Lastly, we have shown that the hybrid entangled state between a qubit
and a binary coherent state can be used as a resource state in quantum teleportation. \\


\begin{thebibliography}{90}
\bibitem{piani} M. Piani, S. Gharibian, G. Adesso, J. Calsamiglia, P.
Horodecki,and A. Winter, Phys. Rev. Lett. \textbf{106}, 220403
(2011);R. Horodecki, P. Horodecki, M. Horodecki, and K. Horodecki,
Rev. Mod. Phys. \textbf{81}, 865 (2009).
\bibitem{Bennett1} C. H. Bennett, G. Brassard, C. Crepeau, R. Jozsa, A. Peres, and W. K. Wootters, Phys. Rev. Lett. \textbf{70}, 1895 (1993).
\bibitem{Bennett2} C. H. Bennett and S. Wiesner, Phys. Rev. Lett. \textbf{69}, 2881 (1992).
\bibitem{hillery} M. Hillery, V. Buzek and A. Berthiaume, Phys. Rev. A
\textbf{59}, 1829 (1999); S. Adhikari, I. Chakrabarty, P. Agrawal,
Quant. Inf. and Comp. \textbf{12}, 0253 (2012).
\bibitem{Gisin} C. H. Bennett, and G. Brassard, in Proceedings of the IEEE International
Conference on Computers, Systems and Signal Processing, Bangalore,
India, (IEEE, New York), 175 (1984); A. K. Ekert, Phys. Rev. Lett.
67, \textbf{661} (1991); N. Gisin, G. Ribordy, W. Tittel, and H.
Zbinden, Rev. Mod. Phys. \textbf{74}, 145 (2002).
\bibitem{schumacher} B. Schumacher, Phys. Rev. A \textbf{54}, 2614 (1996).
\bibitem{fiurasek1}J. Fiurasek, Phys. Rev. A 64, 062310 (2001).
\bibitem{horodecki2} P. Horoecki and A. Ekert, Phys. Rev. Lett. \textbf{89},
127902-1(2002).
\bibitem{fiurasek} J. Fiurasek, Phys. Rev. A \textbf{66}, 052315 (2002).
\bibitem{korbicz} J. K. Korbicz, M. L. Almeida, J. Bae, M. Lewenstein, and A.
Acín, Phys. Rev. A \textbf{78}, 062105 (2008).
\bibitem{bae} J. Bae, Rep. Prog. Phys. \textbf{80}, 104001 (2017).
\bibitem{peres} A. Peres, Phys. Rev. Lett. \textbf{77}, 1413 (1996).
\bibitem{horodecki1} M. Horodecki, P. Horodecki, and R. Horodecki, Phys. Lett. A \textbf{223}, 1 (1996).
\bibitem{keyl} M. Keyl and R. F. Werner, Phys. Rev. A \textbf{64},
052311 (2001).
\bibitem{tanaka} T. Tanaka, Y. Ota, M. Kanazawa, G. Kimura, H. Nakazato, and F.
Nori, Phys. Rev. A \textbf{89}, 012117 (2014).
\bibitem{lim} H-T. Lim, Y-S. Kim, Y-S. Ra, J. Bae, and Y-H.
Kim, Phys. Rev. Lett. \textbf{107}, 160401 (2011).
\bibitem{Verstraete} F. Verstraete, and H. Verschelde, Phys. Rev. Lett. \textbf{90}, 097901 (2003).
\bibitem{kwong} C. J. Kwong, S. Felicetti, L. C. Kwek, J. Bae,
quant-ph/arXiv1606.00427.
\bibitem{horodecki}M. Horodecki, P. Horodecki, and R. Horodecki, Phys. Rev. A \textbf{60},
1888 (1999).
\bibitem{muller}C. R. Muller, G. Leuchs, C. Marquardt, and U. L. Andersen, Phys. Rev. A \textbf{96},
042311 (2017).
\bibitem{yurke}B. Yurke, and D. Stoler, Phys. Rev. Lett. \textbf{57}, 13 (1986).
\bibitem{sanders}B. C. Sanders, and G. J. Milburn, Phys. Rev. A \textbf{45}, 1919 (1992).
\end{thebibliography}
\end{document}